\documentclass[
 aps, pra,
 amsmath,amssymb,
 11pt,
 final,
tightenlines,
 twoside,
 twocolumn,
 nofloats,
nofootinbib,
 superscriptaddress,
showkeys,
showkeywords,
 ]
{revtex4-2}

\usepackage{multirow}
\usepackage{mathtools}
\usepackage[T2A]{fontenc}
\usepackage[utf8x]{inputenc}
\usepackage[english]{babel}
\usepackage{graphicx}         
\usepackage{dcolumn}          
\usepackage{bm}               
\input{maik.rty}
\setcitestyle{authoryear,round}
\def\squareforqed{\hbox{\rlap{$\sqcap$}$\sqcup$}}

\def\sq{\ifmmode\squareforqed\else{\unskip\nobreak\hfil
\penalty50\hskip1em\null\nobreak\hfil\squareforqed
\parfillskip=0pt\finalhyphendemerits=0\endgraf}\fi}

\def\utw{\smash{\rlap{\lower5pt\hbox{$\sim$}}}}

\def\udtw{\smash{\rlap{\lower6pt\hbox{$\approx$}}}}

\def\diameter{{\ifmmode\mathchoice
{\ooalign{\hfil\hbox{$\displaystyle/$}\hfil\crcr
{\hbox{$\displaystyle\mathchar"20D$}}}}
{\ooalign{\hfil\hbox{$\textstyle/$}\hfil\crcr
{\hbox{$\textstyle\mathchar"20D$}}}}
{\ooalign{\hfil\hbox{$\scriptstyle/$}\hfil\crcr
{\hbox{$\scriptstyle\mathchar"20D$}}}}
{\ooalign{\hfil\hbox{$\scriptscriptstyle/$}\hfil\crcr
{\hbox{$\scriptscriptstyle\mathchar"20D$}}}}
\else{\ooalign{\hfil/\hfil\crcr\mathhexbox20D}}%
\fi}}

\begin{document}

\title{Study of the  Cygnus~X-3 Microquasar with  the RATAN-600 Radio telescope \\ in Multi-Azimuth Observing Mode}
\author{\firstname{S.~A.}~\surname{Trushkin}}
\email{sergei.trushkin@gmail.com}
\affiliation{Special Astrophysical Observatory,  Russian Academy of Sciences,
Nizhnii Arkhyz, 369167 Russia}
\affiliation{Kazan (Volga Region) Federal University, Kazan, 420008 Russia}
\author{\firstname{A.~V.}~\surname{Shevchenko}}
\affiliation{Special Astrophysical Observatory,  Russian Academy of Sciences,
Nizhnii Arkhyz, 369167 Russia}
\author{\firstname{N.~N.}~\surname{Bursov}}
\affiliation{Special Astrophysical Observatory,  Russian Academy of Sciences,
Nizhnii Arkhyz, 369167 Russia}
\author{\firstname{P.~G.}~\surname{Tsybulev}}
\affiliation{Special Astrophysical Observatory,  Russian Academy of Sciences,
Nizhnii Arkhyz, 369167 Russia}
\author{\firstname{N.~A.}~\surname{Nizhel'skii}}
\affiliation{Special Astrophysical Observatory,  Russian Academy of Sciences,
Nizhnii Arkhyz, 369167 Russia}
\author{\firstname{A.~N.}~\surname{Borisov}}
\affiliation{Special Astrophysical Observatory,  Russian Academy of Sciences,
Nizhnii Arkhyz, 369167 Russia}
\author{\firstname{A.~A.}~\surname{Kudryashova}}
\affiliation{Special Astrophysical Observatory,  Russian Academy of Sciences,
Nizhnii Arkhyz, 369167 Russia}

  \begin{abstract}
    We have been performing daily observations of bright microquasars at
    1.2--20~GHz with the Northern sector of RATAN-600 radio telescope
    for more than ten years. During the  2019--2021 observations
    we recorded bright flares, which we call giant flares because
    fluxes reach record levels---above 20~Jy---during these events. In
    this paper we report the results of intraday variations of the
    \mbox{Cygnus~X-3} microquasar in multi-azimuth observations made
    with the  ``North sector with a flat-sheet reflector''  during giant
    flares of Cygnus~X-3. These were the first such observations made
    simultaneously at several frequencies on a  short time scale (10
    minutes). Observational data consists of 31 measurement made within
    $\pm$\,2.7~hours of the culmination of the object. We are the first
    to discover the evolution of the spectrum of the flare emission of
    Cygnus~X-3 on a time scale comparable to the orbital period of the
    binary. The measurement data allowed us to determine the temporal
    and spectral parameters of radio emission, which are typical for
    synchrotron flare emission in relativistic jets. Evolution of the
    radio emission of X-ray binaries on short time scales is a key to
    understanding the formation of jet outbursts in the process of mass
    accretion of the matter of the donor star onto the relativistic
    object.
   \end{abstract}
   \maketitle

    \section{INTRODUCTION}
    \begin{figure*}
	\includegraphics[scale=0.5]{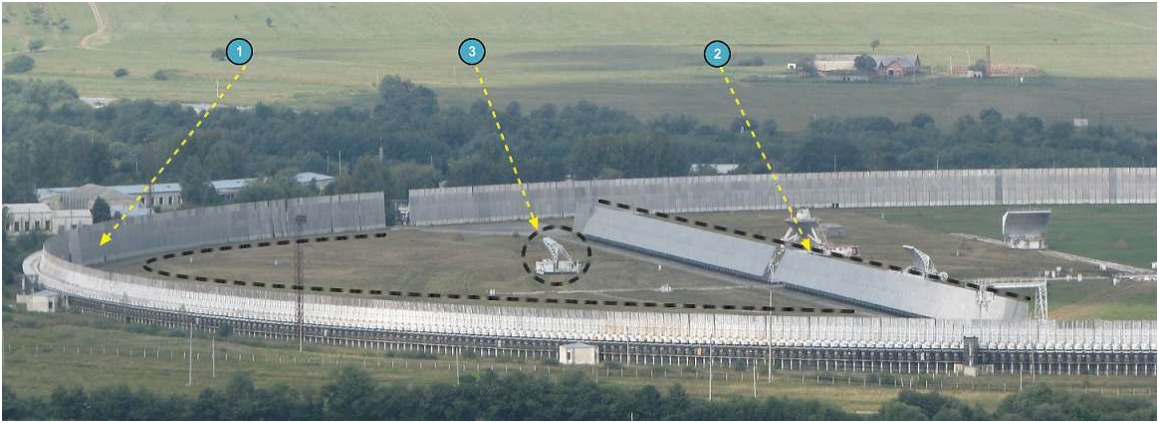}
	\caption{The ``Southern sector with a flat reflector'' antenna system of RATAN-600 radio telescope.
	    (1)---Southern sector of the antenna, (2)---Flat-sheet reflector, (3)---secondary (third) reflector of the feed cabin.}
	\label{fig1}
    \end{figure*}

    The   \mbox{Cygnus~X-3} X-ray binary---a microquasar with relativistic
    jets---consists of a massive Wolf-Rayet star (van Kerkwijk et al.,
    1992) and a compact object (a black hole or a neutron star). The
    orbital period of this system determined from IR and X-ray data is
    4.8~hours (Bhargava et al., 2017), i.e., the system is very compact
    and strong wind from the donor star is one of the key factors
    influencing the features of jet outflows during bright flares.
    The object was found to be an X-ray object in  1967 (Giacconi et al.,
    1967), and powerful radio flares of Cygnus~X-3 have been observed
    since 1972 when a the properties of variable radio emission were
    discussed in detail in a series of 22 papers published in  ``Nature''
    (see Gregory et al. (1972)).  
    long-term monitoring In the \mbox{1980s--1990s} long-term monitoring
    of radio emission was produced with the NRAO interferometer (USA)
    (Waltman et al., 1994, 1996).     After the launch of X-ray space
    observatories with daily monitoring programs (RXTE, BATSE, Swift) it
    became clear that radio flares and X-ray flux are strongly correlated
    in all microquasars  (McCollough et al., 1999).
	This property was named ``coupling'' of different processes in
	X-ray binaries that ensure mass accretion onto the
    relativistic object. This also fully applies to Cygnus~X-3, for
    which the X-ray ``hardness\,--\,intensity'' evolutionary dependence
    resembling, on the whole, similar diagrams for other X-ray binaries
    with accreting black holes (Koljonen et al., 2010).  There is a
    molecular cloud on the Cygnus~X-3 line of sight, which causes strong
    absorption in the visible preventing optical spectroscopy. The object
    is at the distance of 7.2~kpc. Gamma-ray flux of Cygnus~X-3 also
    correlates with the radio flare activity  (Tavani et al., 2009;
    Corbel et al., 2012) and it is studied extensively by  Fermi/LAT
    and AGILE space observatories.
     Zdziarski et al. (2018)  did a detailed analysis of simultaneous
     radio (AMI-LA) and gamma-ray (Fermi/LAT) measurements.
    Both types of measurements were found to exhibit a weak flux
    modulation with the orbital period.  The object has been VLBI mapped
    repeatedly with observations revealing jet outflows with relativistic
    jet motions s (Miller-Jones et al., 2004; Tudose et al., 2007).
     Such motions were in some cases interpreted as ``superluminal'',
    which can most likely be explained as motion in jets directed at
    small angles to the line of sight
     (Spencer, 1998).       It is evident that the time scale of
     variability also shortens in this case.

    Cygnus~X-3 stands out among microquasars by its occasional giant
    radio flares when flux may increase in few days by factors of several
    thousand that occur compared to the preflare level (Waltman et al.,
    1996).  These giant flares occur immediately after the end to the
    state of almost complete ``quenching'', i.e., after the 4--11~GHz
    radio flux has remained below \mbox{5--20}~mJy for one to three
    weeks   (Trushkin et al., 2017).  Such a period of low radio flux
    is also characterized by a hypersoft X-ray state during which hart
    X-ray flux (15--50~keV) drops to zero, whereas the level of soft
    X-ray flux ($<15$~keV) remains high or even increases  (Koljonen
    et al., 2010).  In such a state relativistic radio jets probably
    disappear completely. The intraday variability of Cygnus~X-3 has been
    detected repeatedly, especially at millimeter- and centimeter-wave
    frequencies, which is not surprising given that the compact size of
    the system, relativistic nature of jets, and small absorption of
    radio emission. However, simultaneous multi-frequency measurements
    are of special interest for a detailed comparison and search for
    new patterns.

    \section{MULTI-AZIMUTH OBSERVATION MODE}

    Observations were made with a three-mirror system  ``Southern sector
    with a flat reflector'' (Fig.~1).  This antenna configuration allows a
    cosmic source to be monitored by changing the azimuth of the parabolic
    southern sector within $\pm30^\circ$, and the elevation of the flat
    mirror and azimuth of the third mirror within $\pm30^\circ$.

    The third mirror can relatively rapidly move along arch rails with
    radius of about 150~m, which allows changing the azimuth of the
    antenna focus and thereby track the source in the sky. In 2018 three
    radiometers operating in the 4.7, 8.6, and 16~GHz frequency ranges
    were mounted in the focal plane of the third mirror (see Table~1).

    \renewcommand{\baselinestretch}{0.8}
    \begin{table} []
	\caption {Parameters of radiometer facilities.
	Designations: $\nu$---the central frequency (GHz);
	    $\Delta\nu$---the bandwidth (GHz);
	    $\Delta S_\nu$---the sensitivity in terms of
	    flux per resolution beam  (mJy/beam),
	    and Phi05---the halfwidth of the antenna beam
	    in arcseconds for the transit of a source at the
	    declination of $\delta\sim 42^\circ$ }
	\label{truskin_tab1}
	\medskip
	\begin{tabular}{l|c|c|c}
	    \hline
	    $\nu$,         & $\Delta\nu$,  & $\Delta S_\nu$, & Phi05,  \\
	    GHz & GHz & mJy/beam        & arcsec \\
		    \hline
	    \multicolumn{4}{c}{Northern sector}  \\
	    \cline{1-4}
	    $4.7$     & $0.6$   & $5$     & $  48 $ \\
	    $8.2$     & $1$     & $10$    & $   30$   \\
	    $11.2$    & $1.4$  & $15$   & $   21$ \\
	    $21.7$    & $2.5$   & $50$  & $    15$ \\
	    \hline
	    \multicolumn{4}{c}{Southern sector with a flat reflector }  \\
	    \cline{1-4}
	    $4.7^*$   & $0.6$   & $10$      &  $   72  $   \\
	    $4.7$      &  $0.6$   & $10$      &  $   72  $   \\
	    $8.6$      &  $1.4$   & $20$      &  $   39  $    \\
	    $16$       &  $2$      & $60$      &  $   30  $  \\
	    \hline
	    \multicolumn{4}{l}{\footnotesize {$^*$---four-beam facility}}\\
	\end{tabular}
    \end{table}
    \renewcommand{\baselinestretch}{1.0}

    In 2019 one of the radiometers of the facility of sensitive 4.7~GHz
    four-channel detectors operating in the full-power mode and used
    to search for rapid radio flares was temporarily mounted in the
    focal plane of the third mirror. The broad 600~MHz-wide bandwidth
    of each radiometer was subdivided into four narrow  150-MHz wide
    sub-bands. At the output of each narrow channel ADC were installed
    after the quadratic detector to record digital signals. In these
    measurements signals in the narrow detectors of this radiometer
    were averaged in the process of reduction of observations. This
    detector replaced the previous one at 4.7~GHz because it provided
    higher sensitivity throughout a broader bandwidth.

    The discrete setting of the antenna with a step of $2^\circ$ in
    azimuth made it possible to make 31 measurement of the object,
    which in the case of Cygnus~X-3 was implied  10-minute intervals
    between the measurements.  The apparent positions of the sources
    were computed using  \texttt{EFRAT} ephemeris program developed at
    the Main Astronomical Observatory of the Russian Academy of Sciences
    specifically for the use on RATAN-600 telescope
     (Korzhavin et al., 2012).

     The working  \texttt{EFRAT} program is available from the RATAN-600
     server and
    it is routinely used for solar measurements and for measurements
    of other Solar-system planets.  The basic antenna parameters
    (effective area and beam size) were measured during observations
    of calibrating sources (Fig.~2 and Fig.~3). The antenna efficiency
    decreases appreciably at extreme azimuths because of the reduction
    of the aperture of the Southern sector, which is limited to one
    quarter of the circular reflector of RATAN-600. The beam size along
    the trajectory of the source transit through fixed beam varied
    because of the change of the parallax angle. The vertical beam
    size also varied, from 37$^\prime$ at the meridian to 50$^\prime$
    at $\pm30^\circ$ azimuths.

    We employed the routine RATAN-600 procedure for full-power radiometers
    for daily monitoring of \mbox{Cygnus~X-3} at the meridian, its
    detailed review can be found in  Tsybulev et al.~(2018).
    Low-noise HEMT transistor based  amplifiers are mounted on all
    radiometers.

    \begin{figure}[t]
	\includegraphics[width=0.94\linewidth, bb=1 170 540 580,clip]{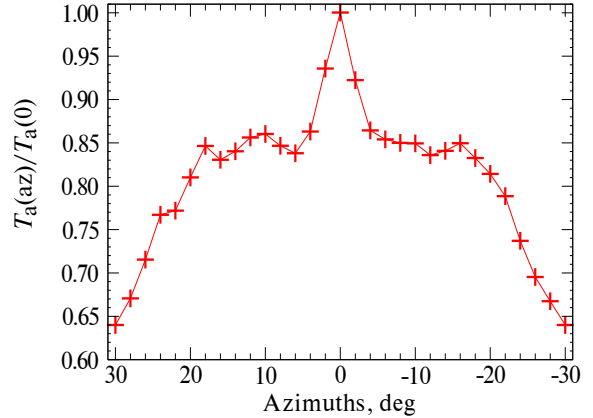}
	\caption{Dependence of normalized antenna temperatures of NGC\,7027 on the azimuth of the  ``Southern sector with a flat
	    reflector'' antenna system. }
	\label{fig2}
    \end{figure}
    \begin{figure}[h!!!] \vspace{1mm}
	\includegraphics[width=0.95\linewidth, bb=10 480 530 820,clip]{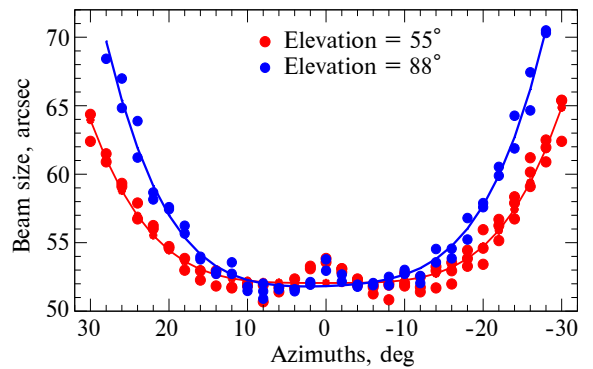}
	\caption{Size of the cross sections of the ``Southern sector with
		 a flat reflector'' multi-azimuth antenna system
		 for two sources at two different declinations.}
	\label{fig3}
    \end{figure}

    Flux calibration in multi-azimuth mode was performed by observing
    the bright radio galaxy \linebreak NGC\,1275 (3C\,84, J0319+41),
    planetary nebula NGC\,7027 (J2107+42), and the H\,II region DR\,21
    (J2039+42). Observations of calibrating sources at close declinations
    were performed similarly to observations of the microquasar studied
    thereby providing accurate flux calibration and control of antenna
    parameters at different azimuths.  The calibrating sources were
    observed several times during the entire cycle of Cygnus~X-3 intraday
    observations.  The radio spectrum of J0319+41 was preliminary
    measured in observations with the Northern sector and the measured
    fluxes were converted into the fluxes at the frequencies used in
    multi-azimuth mode.  We adopted the  NGC\,7027 and DR\,21 fluxes
    in accordance with precision measurements by  Ott et al. (1994)
    taking into account the secular flux decrease due to the expansion of
    the  NGC\,7027 shell.  Our experience with observations of reference
    sources suggests that the total flux calibration error is not greater
    than 3\%.

    We reduced our observational data using
     \texttt{FADPS} software  (Verkhodanov, 1997).
    To increase the  $S/N$ ratio, when reducing the data we convolved
    observational records with the computed beam pattern of
    \mbox{RATAN-600} or with the drift scans of bright reference
    sources.  The accuracy of flux measurements at levels higher than
    1~Jy was better than  3\%  at 4.7~GHz and about  5--10\% at 8.6
    and 16~GHz. The error of the measured spectral index on
    the plots of variation of this parameter was  $\pm\,$0.05 or
    higher. The high measurement accuracy at 4.7~GHz was due
    to high flux values during flares, i.e., the signal exceeded
    noise by several factors of ten.
    That is why the error bars in light-curve plots are smaller
    than the flux value signs.

    Summary Table~2 lists the data of \mbox{Cygnus~X-3} multi-azimuth
    observations and  the reference sources used in observations.
    During the date intervals mentioned in the table the \mbox{Cygnus~X-3}
    monitoring program was carried out daily on the Northern sector.
    \renewcommand{\baselinestretch}{0.7}
    \begin{table*} []
	\caption {Table of  2019--2021 multi-azimuth observations}
	\label{truskin_tab2}
	\medskip
	\begin{tabular}{l|c|c}
	    \hline
	    \multicolumn{1}{c|}{\multirow{2}{*}{Observation dates}}    & \multirow{2}{*}{MJD}             &  Reference   \\
	    &            &  sources   \\
	    \hline
	    April 19--28, 2019  & 58592--58601      &   DR\,21+3C\,84    \\
	    June 6--8, 2019, June  10, 2019     & 58640--58644      &   DR\,21+3C\,84    \\
	    June 18--29, 2019       &  58652--58666     &  DR\,21+3C\,84   \\
	    February 5--9, 2020, February 15--17, 2020  &  58884--58896     &  NGC\,7027         \\
	    December 25--30, 2021    &  59573--58579     &  NGC\,7027         \\
	    \hline
	\end{tabular}
	\renewcommand{\baselinestretch}{1.0}
    \end{table*}
    \begin{figure}[t]
	\includegraphics[width=0.93\linewidth, bb = 10 280 510 643,clip]{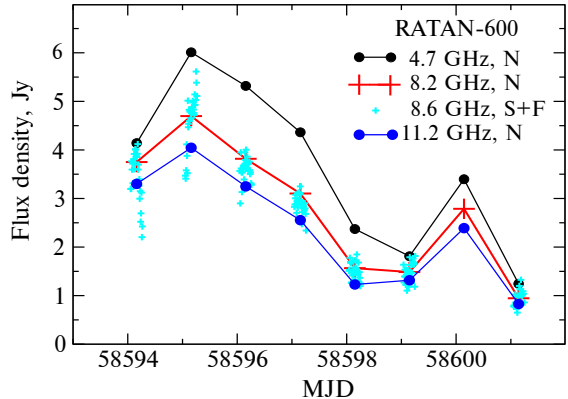}
	\caption{ Daily measurements of Cygnus~X-3 made  in April 2019 with the Northern sector (the black, red, and blue crosses) and
	    Southern sector with a flat reflector (crosses).}
	\label{fig4}
    \end{figure}
    \begin{figure}[]
	\includegraphics[width=0.93\linewidth]{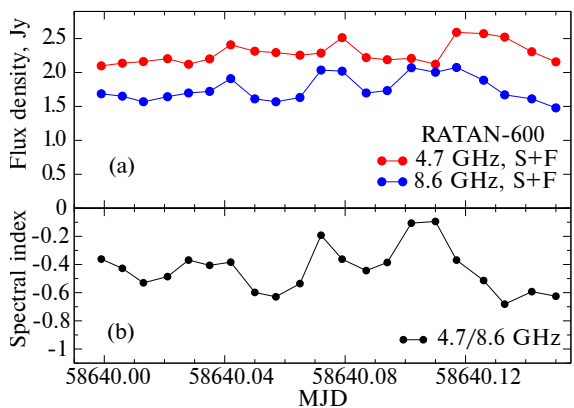}
	\caption{Intraday Cygnus~X-3 variability according to measurements
		 made on June~6, 2019. The bottom panel shows
		 spectral index variations.}
	\label{fig5}
    \end{figure}

    \section{RESULTS OF OBSERVATIONS}

    \subsection{Flare activity of Cygnus~X-3 in April and June, 2019}

    \begin{figure}[]
	\includegraphics[width=0.9\linewidth, bb=10 90 655 590,clip]{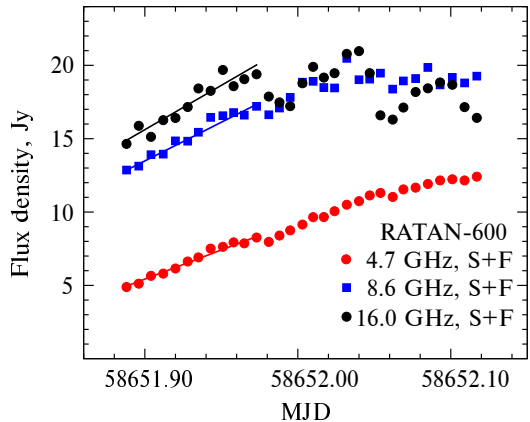}
	\caption{Multi-azimuth flux measurements of the microquasar
		 Cygnus~X-3 made with the Southern sector with a flat reflector
		 during the first day of the flare,
	    on June 18, 2019.}
	\label{fig6}
    \end{figure}
     \begin{figure*}[] \vspace{1mm}
	\includegraphics[width=0.72\linewidth]{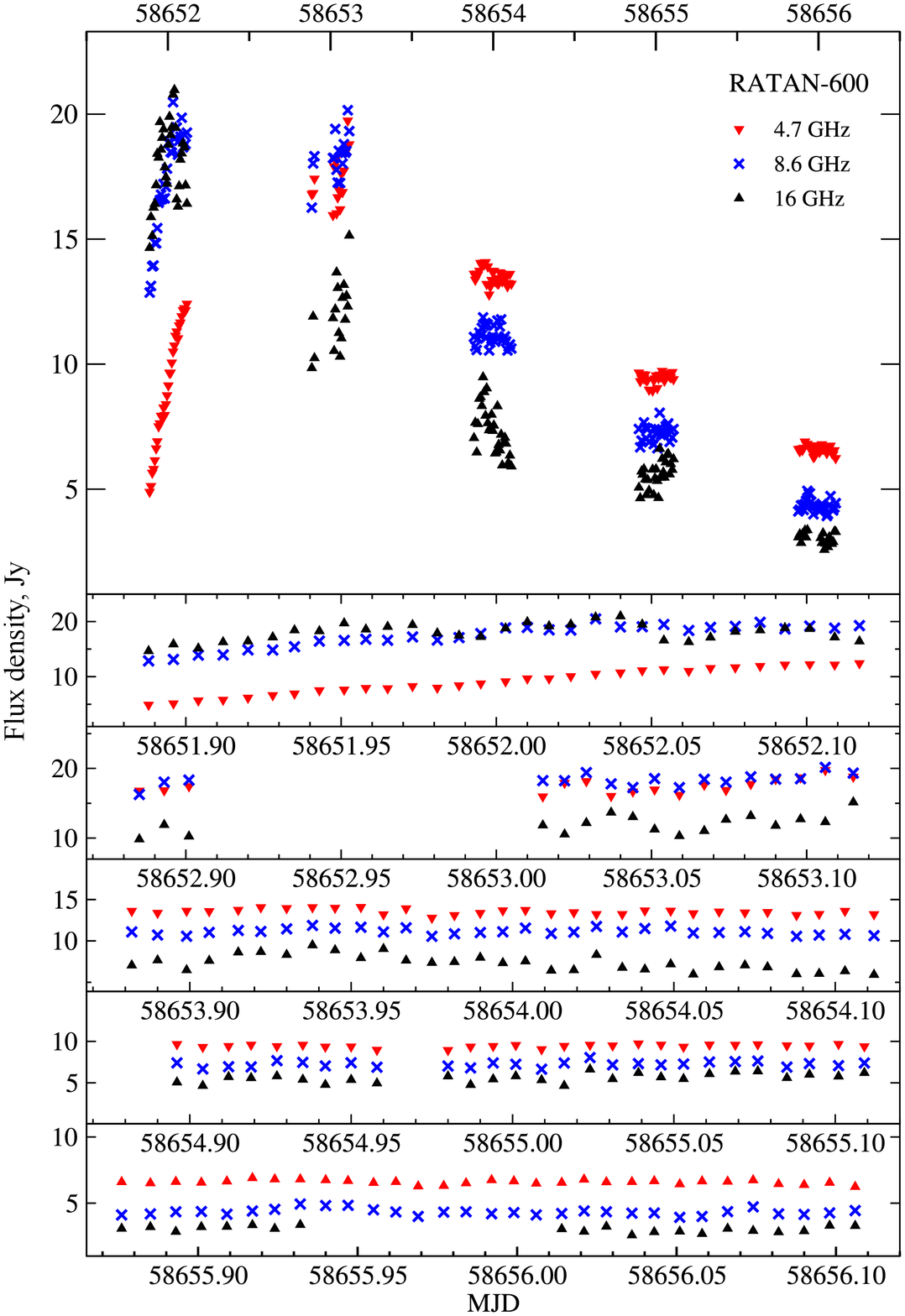}\vspace{-1mm}
	\caption{Cygnus~X-3 light curves at three frequencies during five days: June 18--22,  2019.} \vspace{-1mm}
	\label{fig7}
    \end{figure*}

    This Cygnus~X-3 activity period included two flare states: in
    \mbox{April\,\!--\,\!May}(Trushkin et al., 2019)
     and in   June, 2019  (Trushkin et al., 2020).  In April, 2019 the
     multi-azimuth observation mode was for the first time used
    to study microquasars  (Fig.~4).  In June, 2019 Cygnus~X-3 variability
    monitoring started. In the date intervals listed in Table~2 intraday
    object flux measurements were made during daily  observations at 4.7,
    8.6, and 16~GHz. The resulting light curve are plotted in Figs.~5--7.
    It was during monitoring of 2019 radio flares that rapid (about
    10~minutes) variability of Cygnus~X-3.

    Small flares were observed in early June.  One of these flares
    coincided with a gamma-ray flare recorded by  Fermi/LAT space
    observatory
    \footnote{\url{https://fermi.gsfc.nasa.gov/ssc/data/access/lat/msl_lc/source/Cygnus_X-3}}.
    Figure~5 shows the light curves and variations of spectral index
    during June 6, 2019 (MDJ\,58640). As is evident from the figure,
    during small flares the spectrum was closer to flat, but then
    transformed into the usual optically thin spectrum.

    An important feature of our observations were Cygnus~X-3
    flux measurements at the early stage of the flare evolution.
    This event was recorded on June 18, 2019 and the stage of
    gradual increase of the source flux at all three
    frequencies---4.7, 8.6, and 16~GHz---was observed (Fig.~6).

    \begin{figure}[]
	\includegraphics[width=0.95\linewidth,bb=120 75 850 600,clip]{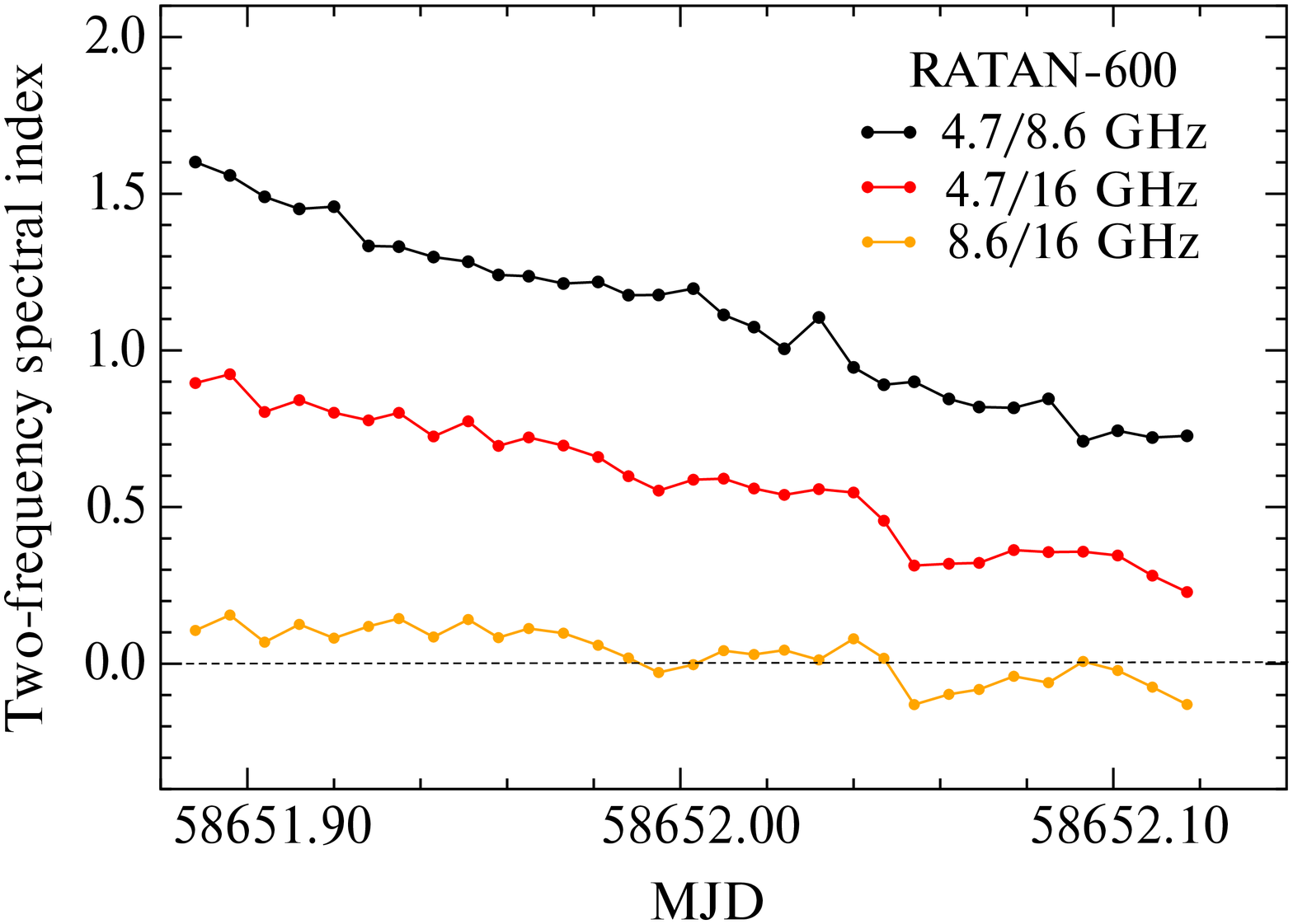}
	\caption{Evolution of the two-frequency spectral index during the
		 June 18, 2019 measurements.}
	\label{fig8}
    \end{figure}

    The flux increase (up to the time of flare maximum) can be closely
    approximated by linear relation: $ S_{\nu} \propto A  ( t-t_0 )$,
    where  $t_0$ is the time of the flare onset.  We extrapolated this
    linear function to determine the ejection starting times 58561.75,
    58561.60, and 58561.50 at 4.7, 8.6, and 16~GHz, respectively. We
    chose the same approximation interval in time for all three
    frequencies. Although the accuracy of such extrapolation is hardly
    better than one hour, it is worth pointing out the appreciable shift
    of the we flare onset time from the high to low frequency, which can
    be easily explained in term of the model of a jet ejection rising
    in an absorbing shell.

    According to the June 18, 2019 measurements, the  $\alpha_{4.7/8.6}$
    spectral index for  \mbox{4.7--8.6~GHz} frequencies changed from
    +1.6 to +0.7, whereas the $\alpha_{8.6/16.0}$ \mbox{(8.6--16.0~GHz)
    spectral index} was close to zero (Fig.~8). This result leads us to
    conclude that radio emission transformed from optically thick into
    optically thin state after the flare reached the maximum brightness
    level. Such spectral evolution is consistent with the radio-jet
    model with synchrotron self absorption operating at the beginning of
    the flare.  Another possible scenario involves absorption by a mix of
    thermal and relativistic electrons in the jet.  Absorption mechanisms
    are difficult to unambiguously separate without using widely separated
    measurement frequencies.

    However, the fact that the spectral index is close to $+2$
    indicates that the thermal absorption mechanism dominates over
    the self-absorption of relativistic electrons.

    \subsection{Flare activity of Cygnus~X-3 in February, 2020}

    During the Januray 16, 2020  (MJD~58864) observations the flux
    decreased down to 5~mJy at 4.7~GHz and then remained very low
    over the next 12 days, as is usually the case during the period
    of the hypersoft X-ray state of the object  (Koljonen et al., 2010).
     The first short-term event occurred on MJD~58876, when the flux
    increased to 100~mJy at 4.7~GHz. The second bright event with a flux
    of about 1~Jy was detected at MJD~58881.3, and
    the spectrum remained optically thick. On February 8, 2020 (MJD~58887.3)
    a giant flare started (Fig.~9).
    At that very time simultaneous observations were produced with AMI-LA
    radio telescope (Green and Elwood, 2020; Spencer et al., 2022).
    \begin{figure}[]
	\includegraphics[width=0.9\linewidth, bb=170 50 845 600,clip]{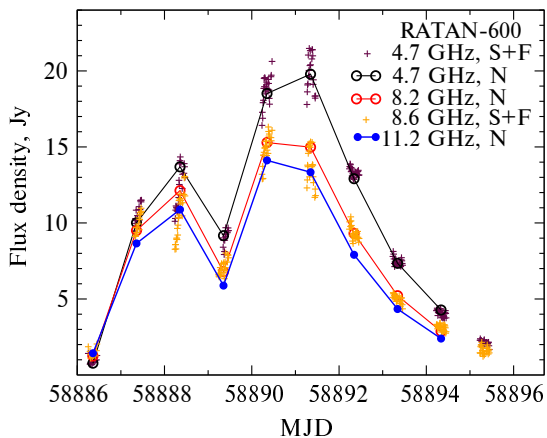}
	\caption{Daily measurements of Cygnus~X-3 made in February, 2000
		 with the Northern sector with a flat reflector (crosses). }
	\label{fig9}
    \end{figure}

    Figure~10 shows the light curves and evolution of the two-frequency
    spectral index during MJD~58888. The flux increase during
    this optically thin flare was accompanied by the flattening of the
    spectrum typical for a flare onset, i.e., the spectral index
    decreased from  $-$0.5 to $-$0.1. Daily multi-frequency monitoring of
    microquasars and intraday measurements during active states of
    Cygnus~X-3 provide further insight into the relationship between the jet and
   the accretion disk in a micriquasar. Synchrotron emission models allow
   estimating the size and brightness temperature of emission regions.
    \begin{figure}[]
	\includegraphics[width=0.9\linewidth, bb=0 0 715 535,clip]{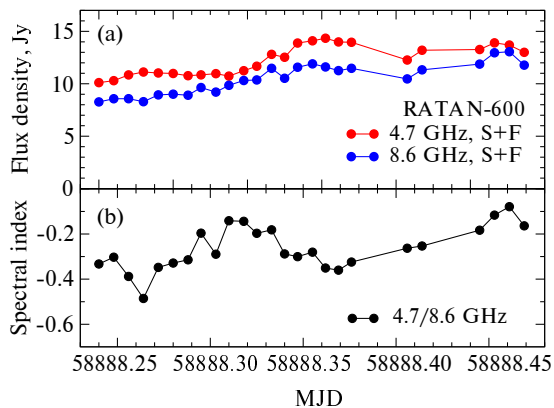}
	\caption{Variability of Cygnus~X-3 on February 9, 2000 and evolution of spectral index.}
	\label{fig10}
    \end{figure}
    \begin{figure*}\vspace{1mm}
	\includegraphics[width=0.77\linewidth, bb=0 0 860 600,clip]{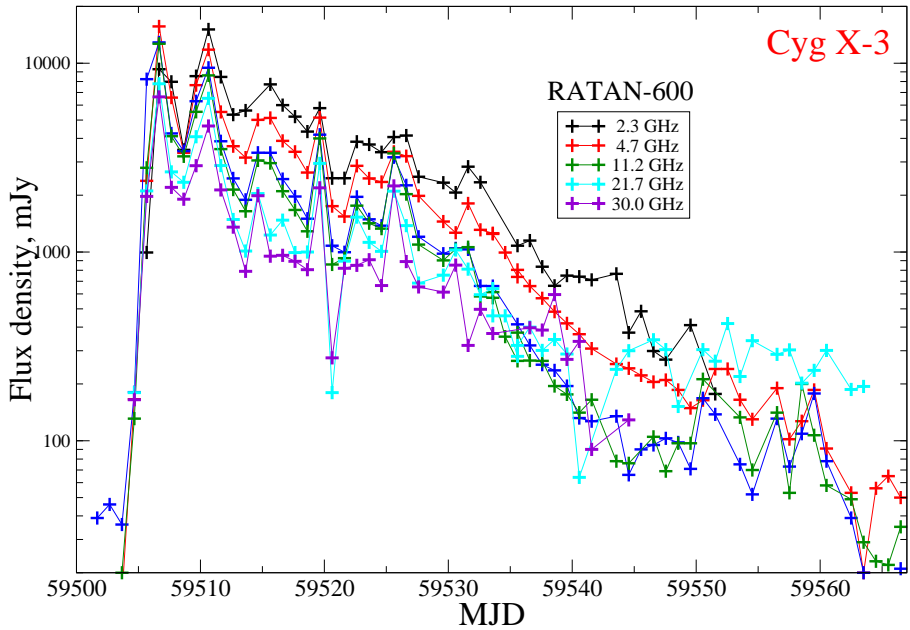}
	\caption{Light curves of Cygnus~X-3 in October--November, 2021 based
		 on daily monitoring data obtained with the Northern sector
	    of RATAN-600 radio telescope. }
	\label{fig11}
    \end{figure*}

    \subsection{Flare activity of Cygnus~X-3 in 2021}
    During the monitoring on July 31, 2021 we detected a giant flare
    from the microquasar Cygnus~X-3 in the  \mbox{1.2--30}~GHz frequency
    interval. This flare was probably characterized by a single ejection
    and lasted for about two weeks.
    Like in most of the cases, the flare was accompanied by
    appreciable increase of $\gamma$-ray flux as it follows from
    the \mbox{0.1--300}~GeV monitoring data for the object
    obtained from  Fermi/LAT space observatory in MJD~59425.0.
    The gamma-ray event was recorded a little less than one
    day before we detected the 4.7~GHz radio flux maximum
    at MJD~59425.88.  According to \mbox{15--50}~keV  Swift/BAT
    data\footnote{\url{https://swift.gsfc.nasa.gov/results/transients/CygX-3/}},
    the active X-ray state continued and on October 18, 2021  (MJD~59505)
    we detected an unusual giant positive flare with the 4.7-GHz flux
    at maximum reaching almost 16~Jy (Fig.~11). This flare consisted of
    several consecutive ejections and ended after 45 days, on December 2
    (MJD~59550).

    According to Fermi observatory data, the October flare was also
    accompanied with $\gamma$-ray activity and lasted until
    November 13 (MJD~59531). On December 24, 2021 (MJD~59572.48) we
    detected a giant flare (Fig.~12), which was quite
    predictable because the microquasar remained in hypersoft state.
	And again, such spectrum is determined by optically thick
	(spectral index $\alpha = 1.6$) and optically  thin
    ($\alpha = 0$) modes of the forming jets.
    \begin{figure*}
	\includegraphics[width=0.74\linewidth,bb=0 0 860 600,clip]{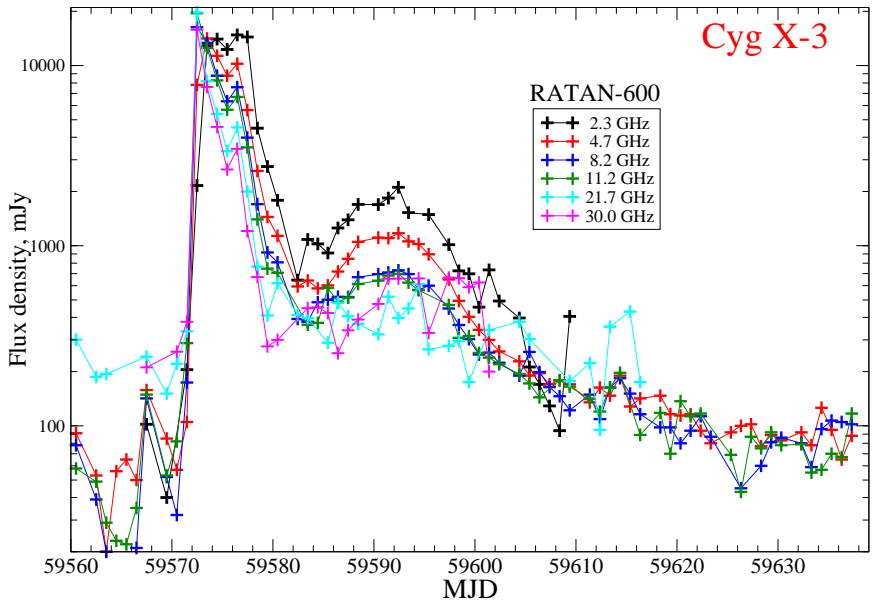}
	\caption{The Cygnus~X-3 light curve at the beginning of a flare in December, 2021 according to monitoring data and
	    multi-azimuth measurements. }
	\label{fig12}
    \end{figure*}
    The \mbox{Cygnus~X-3} then evolved into optically thin state with
    the usual spectral index \mbox{$\alpha = -0.55$}.
    We carried out four sets of observations in azimuths and found only
    small intraday variations of oscillations at 4.7~GHz (within 10\%)(Fig.~13).
    \begin{figure}
	\includegraphics[width=0.92\linewidth, bb=0 0 860 600,clip]{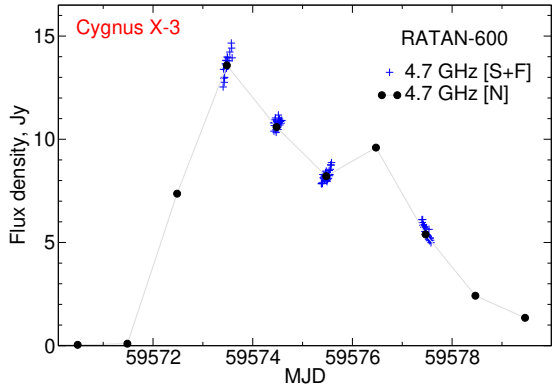}
	\caption{The 4.7-GHz Cygnus~X-3 light curves according to measurements
		 made with the Northern sector (the black dots)
		 and Southern sector with a flat reflector in December, 2021. }
	\label{fig13}
    \end{figure}

    \section{DISCUSSION}

    Radio flares are a good indicator of the formation process of a
    jet ejection and overall increase of activity in the Cygnus~X-3
    microquasar. The onset of the flare at the stage of rapid flux
    increase is of special interest because it provides an insight into
    how electrons responsible for synchrotron radio emission generated.
    In the case of the giant flare of June 18, 2019 we convincingly
    demonstrated that the 4.7, 8.6, and 16~GHz fluxes of the flares
    increased linearly until reaching maximum levels and with a small
    delay for low frequencies.

    Note that rapid evolution of optically thick radio emission imposes
    significant constraints on the jet geometry.  Indeed, after the
    flare onset volume clumps inside jets must usually expand in three
    dimensions in accordance with the classical model  van der Laan
    (1966).  However, such expansion may be exponential for jets with
    conic geometry (Marti et al., 1992), when rather thin shock
    envelopes in jets may result in two-dimensional expansion.

    The measured spectra were indicative of optically thick emission,
    which was probably associated with internal absorption by thermal
    electrons or with synchrotron self-absorption in accordance with
    Fender and Bright (2019).  Smooth intraday radio flux variations
    was also detected at the stage of flare decay.

    Here we focused only on the results of observations in the new mode
    for the context considered. We defer a more detailed analysis and
    the application of various models to the accompanying paper.

    \section*{ACKNOWLENGMENTS}
      Observations were performed with the equipment of RATAN-600 radio
      telescope of the Special Astrophysical Observatory of the Russian
      Academy of Sciences and supported by the Ministry of Science and
      Higher Education of the Russian Federation.  We are sincerely
      grateful to the two reviewers for their constructive comments
      that helped to improve the paper.

    \section*{FUNDING}
    Part of the observational data was exposured on the unique scientific
    facility  the radio telescope RATAN-600
    SAO RAS and the  data processing  was supported  under  the   Ministry
    of Science and Higher Education of the Russian Federation grant
   \mbox{No.~075-15-2022-262} (13.MNPMU.21.0003).

    \section*{CONFLICT OF INTEREST}
    The authors declare that there is no conflict of interest.

\end{document}